\documentclass[preprint,10pt]{sigplanconf}

\usepackage{amsmath}

\usepackage{pstricks}
\usepackage{xspace}
\usepackage{graphicx}

\usepackage{bold-extra}
\usepackage{listings}
\usepackage{fancybox}

\usepackage{url}


\lstdefinelanguage{JavaScript}{
  keywords={break, case, catch, continue, debugger, default, delete, do, else, finally, for, function, if, in, instanceof, new, return, switch, this, throw, try, typeof, var, void, while, with},
  morecomment=[l]{//},
  morecomment=[s]{/*}{*/},
  morestring=[b]',
  morestring=[b]",
  sensitive=true
}

\newcommand{\TODO}[1]{}
\renewcommand{\TODO}[1]{{\bfseries\color{red} TODO: {#1}}}

\begin{document}

\setlength{\pdfpageheight}{\paperheight}
\setlength{\pdfpagewidth}{\paperwidth}

\conferenceinfo{CONF 'yy}{Month d--d, 20yy, City, ST, Country}
\copyrightyear{20yy}
\copyrightdata{978-1-nnnn-nnnn-n/yy/mm} 
\doi{nnnnnnn.nnnnnnn}

\title{Interprocedural Type Specialization of JavaScript Programs Without Type Analysis}

\authorinfo{Maxime Chevalier-Boisvert}
           {DIRO, Universit\'e de Montr\'eal, Quebec, Canada}
           {chevalma@iro.umontreal.ca}
\authorinfo{Marc Feeley}
           {DIRO, Universit\'e de Montr\'eal, Quebec, Canada}
           {feeley@iro.umontreal.ca}

\maketitle



\begin{abstract}
Dynamically typed programming languages such as Python and JavaScript defer
type checking to run time.
VM implementations can improve performance by eliminating redundant dynamic
type checks. However, type inference analyses are often costly and involve
tradeoffs between compilation time and resulting precision. This has lead to
the creation of increasingly complex multi-tiered VM architectures.

{\em Lazy basic block versioning} is a simple JIT compilation
technique which effectively removes redundant type checks from critical code
paths. This novel approach lazily generates type-specialized versions of basic blocks
on-the-fly while propagating context-dependent type information. This approach
does not require the use of costly program analyses, is not restricted by the
precision limitations of traditional type analyses.

This paper extends lazy basic block versioning to propagate type information interprocedurally, across function call
boundaries. Our implementation in a JavaScript JIT compiler shows that across 26\unskip benchmarks, interprocedural basic block
versioning eliminates more type tag tests on average than what is achievable with static
type analysis without resorting to code transformations.
On average,
94.3\unskip\% of type tag tests are
eliminated, yielding speedups of up to 56\unskip\%.
We also show that our implementation is able to outperform Truffle/JS on
several benchmarks, both in terms of execution time and compilation time.

\end{abstract}

\section{Introduction}\label{sec:intro}



The highly dynamic semantics of JavaScript (JS) make optimization difficult.
Late binding, dynamic code loading and the eval construct make type analysis a
hard problem. Precise type analyses also tend to be expensive, and are
often considered too costly to be used in Just-In-Time (JIT) compilers.

Lazy Basic Block Versioning (BBV)~\cite{bbv_ecoop} is an intraprocedural JIT
compilation strategy which allows rapid and effective generation of
type-specialized machine code on-the-fly without a separate type analysis
pass~(Section~\ref{sec:bbv}).

In this paper we introduce an interprocedural variant which extends BBV
with mechanisms to
propagate type information interprocedurally, across function call boundaries
(Section~\ref{sec:interp}), both function parameter and return value types.
Combining these new elements with BBV yields a lightweight approach to interprocedurally
type-specialize programs on-the-fly without performing global type inference or type
analysis in a separate pass.

A detailed evaluation of the performance implications is provided in
Section~\ref{sec:evaluation}.
Empirical results across 26\unskip benchmarks show that, on average, the
interprocedural BBV approach, as far as eliminating type tests goes, performs better
than a simulated perfect static type analysis (Section~\ref{sec:type_tests}). On average,
94.3\unskip\% of dynamic type tests are
eliminated. Speedups of up to
56\unskip\% are also obtained, with only a
modest increase in compilation time.

\section{Background}


The work presented in this paper is implemented in a research
Virtual Machine (VM) for JavaScript (ECMAScript 5) known as
Higgs\footnote{https://github.com/higgsjs}.
The Higgs VM features a JIT compiler built upon an experimental design centered
around BBV.
This compiler is intended to be lightweight with a simple
implementation. Code generation and type specialization are performed in a
single pass. Register allocation is done using a greedy register allocator.
The runtime and standard libraries are self-hosted, written in an
extended dialect of ECMAScript with low-level primitives. These low-level
primitives are special instructions which allow expressing type tests as well
as integer and floating point machine instructions in the source language.

\subsection{Value Types}

Higgs segregates values into categories based on
type tags~\cite{type_tags}. These type tags form a simple, first-degree notion of types that is
used to drive code versioning.  The {\tt unknown} type is also used by the
code versioning process to indicate that the type is unknown.
These type identifiers are listed in the following table.

\vspace*{1ex}

\begin{center}
\begin{tabular}{|l|l|}
\hline
{\tt int32} & 32-bit integers \\ \hline 
{\tt float64} & 64-bit floating point values \\ \hline
{\tt null} & JS {\tt null} \\ \hline
{\tt const} & miscellaneous JS constants \\ \hline
{\tt string} & JS strings \\ \hline
{\tt array} & JS arrays \\ \hline
{\tt closure} & JS function objects \\ \hline
{\tt object} & other JS objects \\ \hline
{\tt unknown} & type is unknown \\ \hline
\end{tabular}
\end{center}

\subsection{Lazy Basic Block Versioning\label{sec:bbv}}

BBV is a JIT code generation technique originally
applied to JavaScript by Chevalier-Boisvert \&
Feeley~\cite{bbv_ecoop}, and adapted to Scheme by Saleil \&
Feeley~\cite{vers_scheme}. The technique bears similarities to HHVM's
tracelet-based compilation approach and Psyco's JIT code specialization
system~\cite{psyco}.

Consider the recursive JavaScript function {\tt f}
shown in Figure~\ref{fig:f}.  Given that {\tt f}'s parameter {\tt n} is not known to be an {\tt int32}, {\tt float64}, or other type, a non-optimizing JIT compiler would put several
type tests in the generated code, namely for the operators {\tt ==} (type of {\tt n}), {\tt -} (type of {\tt n}), and {\tt +} (type of {\tt n} and type of {\tt f(n-1)}).  With intraprocedural BBV a JIT compiler would determine the type of {\tt n} at the first type test (in {\tt n==0}), and, assuming it is an {\tt int32}, would generate code for the expression {\tt n+f(n-1)} specialized to {\tt n} being an {\tt int32}, thus avoiding the two subsequent type tests of {\tt n}.

\begin{figure}[t]
\begin{lstlisting}[language=javascript]
function f(n) {
    if (n == 0)
        return 0;
    else
        return n + f(n-1);
}
\end{lstlisting}
\caption{Simple example of lazy basic block versioning.\label{fig:f}}
\end{figure}

Note that intraprocedural BBV does not propagate type information between caller and callee functions and so two type tests remain: for the expression {\tt n==0} because the type of {\tt f}'s parameter is unknown, and for the addition, because {\tt f}'s return value type is unknown. This is the motivation for our work.

The BBV approach efficiently generates type-specialized machine code in a single pass,
without the use of costly type inference analyses or profiling. 
This is achieved by lazily cloning and type-specializing single-entry
single-exit basic blocks on-the-fly. As in Psyco, code generation and code
execution are interleaved so that run-time type information can be extracted
by the code generation engine. The accumulated information allows the removal
of redundant type tests, particularly in performance-critical paths.

Type information is accumulated as blocks and type tests are
compiled, and propagated forward to successor blocks. A mapping of live
variable types to specialized block versions is maintained. The technique is
a form of forward type propagation. It is unlike traditional fixed
point type analyses in that instead of computing a fixed point on value types,
it is lazily computing a fixed point on the creation of basic block versions. The
presentation in~\cite{bbv_ecoop} explains strategies to effectively limit
the generation of block versions and prevent a code size explosion.  In practice
a hard limit of 5 versions per basic block yields good results.

Because of its JIT nature, BBV has at least
two powerful advantages over traditional static type analyses. The first is
that BBV focuses on the parts of the control flow graph that
get executed, and it knows precisely which they are, as versions are only
generated for executed basic blocks.
The second is that code paths can often be duplicated and specialized
based on different type combinations, making it possible to avoid the loss of
precision caused by control flow merges in traditional type analyses.

\subsection{Typed Object Shapes}\label{sec:typed-shapes}

BBV, as presented in~\cite{bbv_ecoop}, deals with function parameter
and local variable types, and it has no mechanism for attaching types
to object properties.  This is problematic because, in JS, functions are
typically stored in objects.  This includes object methods and also
global functions (JS stores global functions as properties of
the \emph{global object}).  To allow interprocedural type propagation
it is essential to know which function is being called for as many
call sites as possible, both for calls to global functions and method
calls.

Currently, all commercial JS engines have a notion of object shapes, which is
similar to the notion of property maps invented for the Self VM.
That is, any given object contains a pointer to a shape descriptor providing its
memory layout: the properties it contains, the memory offset each property is
stored at, as well as attribute flags (i.e. writable, enumerable, etc.).
Work done on the Truffle Object Model~\cite{truffle_obj} describes how object
shapes can be straightforwardly extended to also encode type tags for object
properties, so long as property writes are guarded to update object shapes when
a property type changes.

Chevalier-Boisvert and Feeley have extended upon the original BBV work with
typed object shapes~\cite{typed_shapes}, giving BBV the ability to propagate
information about the shape of objects, and also their property types.
Polymorphic Inline Caches (PICs) are broken down into cascades
of shape tests exposed as multiple basic blocks in the compiler Intermediate
Representation (IR). This makes it
possible for the BBV engine to extract and propagate shape
information at every property read and write. Propagating said information then
allows extracting property types from object shapes at property reads, but
also to eliminate redundant PICs after successive property reads and writes.

\begin{figure}[tb]
\begin{lstlisting}[language=javascript]
function Accum() {
    this.n = 0;
    this.add = function id1(x) { this.n += x };
    this.sub = function id2(x) { this.n -= x };
}

var a = new Accum();
a.add(5); 
\end{lstlisting}
\caption{Accumulator object with two methods.\label{fig:accum}}
\end{figure}

\begin{figure}[tb]
\begin{center}
\includegraphics[scale=0.5]{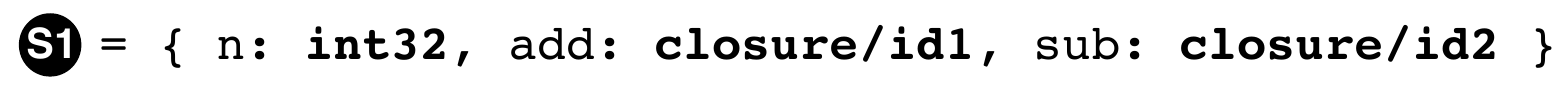}
\end{center}
\caption{Shape of object {\tt a}, encoding the identity of both of its methods.\label{fig:S1}}
\end{figure}

An interesting advantage of typed shapes is that the identity
of methods, when known, is also encoded in object shapes. In practice, this makes it
possible to determine the identity of callees in the large majority of cases. 
This is done by extending the type {\tt closure} so that it carries
the identity of the method.

Consider, for instance, the JS program given in Figure~\ref{fig:accum}. The
identity of the methods assigned in the {\tt Accum} function are encoded in
the shape of the newly constructed object and propagated
to the variable {\tt a}. Hence, in the expression {\tt a.add}, the variable
{\tt a} is known to have the shape given in Figure~\ref{fig:S1}, and the
method call to {\tt a.add} is known to refer to the {\tt id1} method at
code generation time.

In Higgs, method identity is encoded
as a unique pointer to the IR node of that method. It is denoted
{\tt closure/id}$N$ in the examples.  There is also the type
{\tt closure/unknown} needed to handle the situation where it is known
that a value is a function, but the identity of the function is unknown
(this gives more precision than using the {\tt unknown} type, which is
used when nothing is known of the type).
When writing a value of type $T$ to property $P$ of an object with shape $S$,
if $T$ is different than $S.P$ (the type of $P$ in the shape $S$),
the compiler will generate code to create and store in the object a
new shape $S'$ identical to $S$ except that $S'.P=T$.

Method identity is valuable information
because it makes it possible to directly jump to a
callee's entry point without dynamic dispatch. The usefulness of this in the
implementation of interprocedural BBV is explained further
in Section~\ref{sec:entry_point_spec}.

\section{Interprocedural Lazy BBV}\label{sec:interp}
Here we present the main contributions of this paper:
entry point specialization and call continuation specialization.

\subsection{Entry Point Specialization}\label{sec:entry_point_spec}

Procedure cloning has been shown to be a viable optimization technique, both
in ahead of time and JIT compilation contexts. By specializing function bodies
based on argument types at call sites, it becomes possible to infer the types
of a large proportion of local variables, allowing effective elimination of
type checks.

Our first extension to BBV is to allow functions to have multiple type-specialized entry points.
That is, when the identity of a callee at a given call site is known at
compilation time, the JIT compiler requests a specialized
entry point for the callee which assumes the argument types known at the
call site. Type information is thus propagated from the caller to the callee.

Inside the callee, BBV proceeds as described
in~\cite{bbv_ecoop}, deducing local variable types and eliminating redundant
type checks. Our approach places a hard limit on the number of versions that
may be created for a given basic block, and so on the number of entry points
that may be created for any given function. If there are already too many
specialized entry points for a given callee a generic entry point may be obtained instead. This does not matter to
the caller and occurs rarely in practice.

Propagating types from callers to callees allows eliminating redundant
type tests in the callee, but also makes it possible to pass arguments without
boxing them, thereby reducing the overhead incurred by function calls. Note
that our approach does not use any dynamic dispatch to propagate
type information from callers to callees. It relies on information obtained
from typed shapes to give us the identity of callees (both global functions
and object methods) for free. With the current system,
the identity of callees is known at compilation time
in 97\unskip\% of cases on average.
When the identity of a callee is unknown, a generic entry point is used.


\subsection{Call Continuation Specialization}\label{sec:cont_spec}

Achieving full interprocedural type propagation demands
passing the result type information from callees to callers. While it is fairly straightforward to establish
the identity of the single callee a given call site will jump to in the majority of cases, the
points at which a function returns to callers are likely to return
to multiple call continuations within a program. These continuations may in
turn receive input from multiple return points.

Type information about return values could be propagated with a
dynamic dispatch of the return address indexed with the result type.
However this would incur a run time cost.  Instead, our second extension
to BBV uses an approach with a zero run time cost amortized overhead.  The call
continuations are compiled lazily when the first return to a given
continuation is executed. Each time a return statement is compiled,
the type that is being returned is memorized (if known at
compilation time).

When compiling a call continuation, the identity of the callee is checked.
If the identity of the callee is known and its
return type has been determined (so far), the code is specialized based on
this type. If the callee, later during its execution, returns an incompatible
type with the one seen so far, then all call sites of this function
are notified that their call continuations are now invalid and must be
recompiled. Continuations which are invalidated have their code replaced
with a stub that will lazily trigger recompilation with an {\tt unknown}
result type if the continuation is ever executed again.

\section{Extended Example}
\begin{figure}[t]
\begin{lstlisting}[language=javascript]
function sum(tree) {
    if (tree == null)
        return 0;
    else
        return sum(tree.left) +
               sum(tree.right) +
               t.val;
}

function makeTree(depth) {
    if (depth == 0)
        return null;
    else
        return { val: depth,
                 left: makeTree(depth-1),
                 right: makeTree(depth-1)
               };
}

var root = makeTree(8);
if (sum(root) != 502)
    throw Error('error');
\end{lstlisting}
\caption{Binary tree traversal example.\label{fig:tree-sum}}
\end{figure}

For illustration purposes, Figure~\ref{fig:tree-sum} shows the {\tt sum}
function for traversing a balanced binary tree and computing the sum of
numerical values stored in each node. Also shown are functions used to
construct such a tree and test the behavior of the code. While this example
may appear simple, there is much semantic
complexity hiding behind the scene. A correct but naive implementation of this
function contains many implicit dynamic tests.

\begin{figure}[t]
\begin{center}
\includegraphics[scale=0.5]{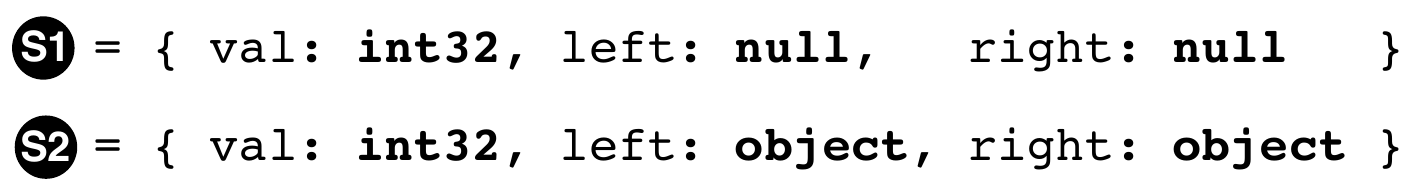}
\end{center}
\caption{Typed shapes for tree nodes seen by the {\tt sum} function\label{fig:S1-S2}}
\end{figure}

Since our system uses typed shapes to specialize code based on object property
types (see Section~\ref{sec:typed-shapes}), the tree nodes produced by
{\tt makeTree} come in two flavors, shown in Figure~\ref{fig:S1-S2}.
Shape {\tt S1} corresponds to leaf nodes, and encodes that both {\tt left} and
{\tt right} node pointers are set to null. 
Shape {\tt S2} corresponds to internal nodes, and encodes that both node
pointers are object references.

\begin{figure*}[tb]
\begin{center}
\includegraphics[scale=0.40]{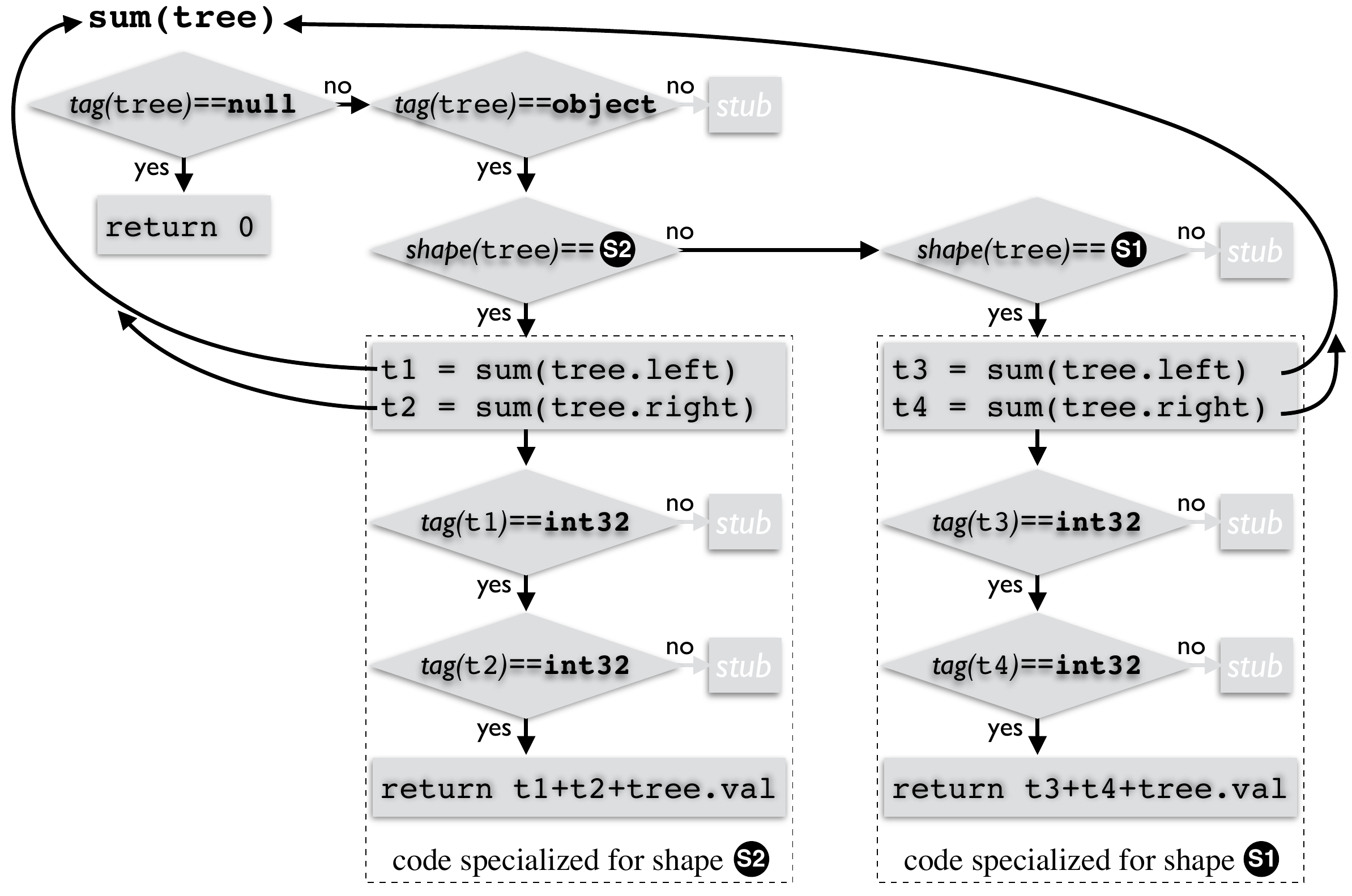}
\end{center}
\caption{Generated code for the {\tt sum} function with intraprocedural BBV\label{fig:intra-code}}
\end{figure*}

Figure~\ref{fig:intra-code} shows a diagram of the code generated for the
{\tt sum} function by Higgs with intraprocedural BBV.
Some details, such as the details of global function lookups and
overflow handling, were omitted from the figure for brevity.
When entering the function, the type tag of {\tt tree} must be
tested. Either the value is {\tt null} or
it is an {\tt object}.
Once {\tt tree} is known to be an {\tt object}, its shape must be tested so that its
{\tt val}, {\tt left} and {\tt right} properties may be accessed. This is
necessary because shapes encode the offsets in memory at which the properties
are stored.  The test for shape S2 is performed first because it is the shape
of {\tt tree} when that code was generated by BBV.

Whether {\tt tree} is a leaf (shape S1) or internal node (shape S2), the code executed is abstractly the same.
Two recursive calls to {\tt sum} are made so that the sums can be computed
for the {\tt left} and {\tt right} children of {\tt tree}.
Higgs knows that {\tt tree.val} is an integer value because that information
is encoded in {\tt S1} and {\tt S2}, but with intraprocedural BBV, does not know what the return types of the recursive calls
to {\tt sum} are. Hence, it must test that {\tt t1} and
{\tt t2} are integers before adding them to {\tt tree.val} with
integer additions.

\begin{figure*}[tb]
\begin{center}
\includegraphics[scale=0.40]{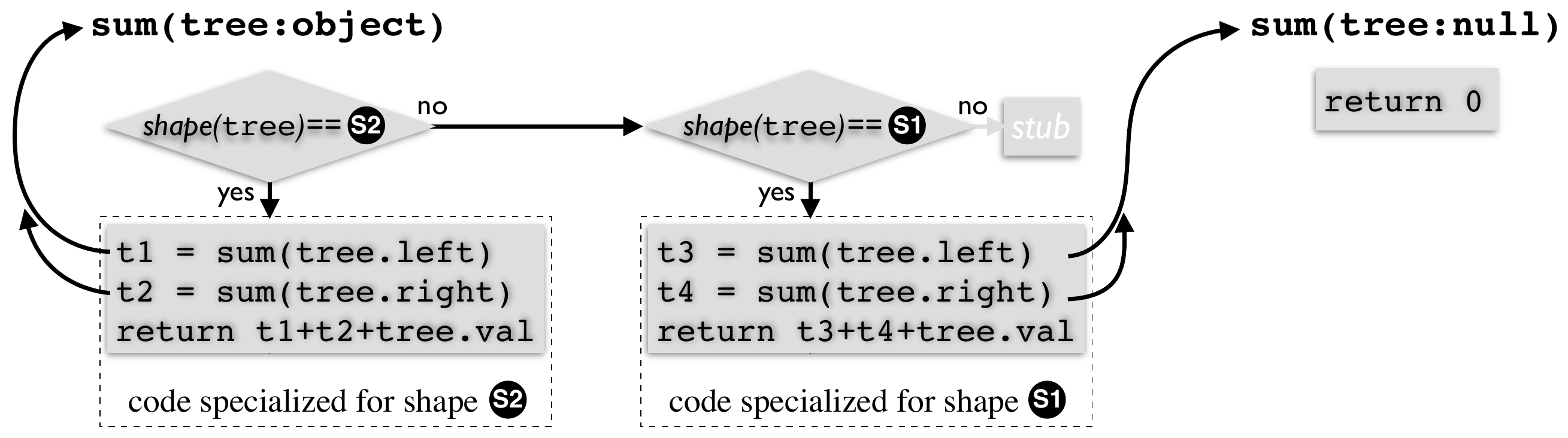}
\end{center}
\caption{Generated code for the {\tt sum} function with interprocedural BBV\label{fig:inter-code}}
\end{figure*}

Figure~\ref{fig:inter-code} shows the code generated for the {\tt sum}
function with entry point versioning and continuation specialization enabled.
Entry point versioning has created
two versions of the {\tt sum} function: one for when the tree is an {\tt object} and one for when the tree is {\tt null}. The version of {\tt sum} specialized for null is
trivial and returns immediately. The version specialized for objects
does two shape checks, as before, but contains no type tag checks.

The elimination of type tag checks is possible because, on entry to
{\tt sum(tree:object)}, {\tt tree} is known to be an object due to entry point
specialization. Furthermore, continuation
specialization determines that {\tt sum} returns {\tt int32} values, which
eliminates the type tag checks on {\tt t1}, {\tt t2}, {\tt t3}, and {\tt t4}. The resulting
code is faster and more compact.

\section{Evaluation}\label{sec:evaluation}
An implementation of the Higgs JIT compiler extended with
entry point and continuation specialization was tested on a
total of 26\unskip classic benchmarks from the SunSpider and V8
suites. One benchmark from the SunSpider suite and one from the V8 suite were
not included in our tests because Higgs does not yet implement the required
features. Benchmarks making use of regular expressions were
discarded because Higgs and Truffle/JS~\cite{trufflejs, truffle} do not
implement JIT compilation of regular expressions.

To measure steady state execution time separately from compilation time in
a manner compatible with both Higgs and Truffle/JS, the benchmarks were
modified so that they could be run in a loop. A number of warmup
iterations are first performed so as to trigger JIT compilation and
optimization of code before timing runs take place.

The number of warmup and timing iterations were scaled so that short-running
benchmarks would execute for at least 1000ms in total during both warmup and
timing. Unless otherwise specified, all benchmarks were run for at least 10
warmup iterations and 10 timing iterations.

Truffle/JS v0.5 was used for performance comparisons. Tests were executed on a
system equipped with an Intel Core i7-4771 CPU and 16GB of RAM running Ubuntu
Linux 12.04. Dynamic CPU frequency scaling was disabled to ensure reliable
timing measurements.

\subsection{Dynamic Type Tests}\label{sec:type_tests}

\begin{figure*}[tb]
    \begin{center}
    \includegraphics[scale=1.00]{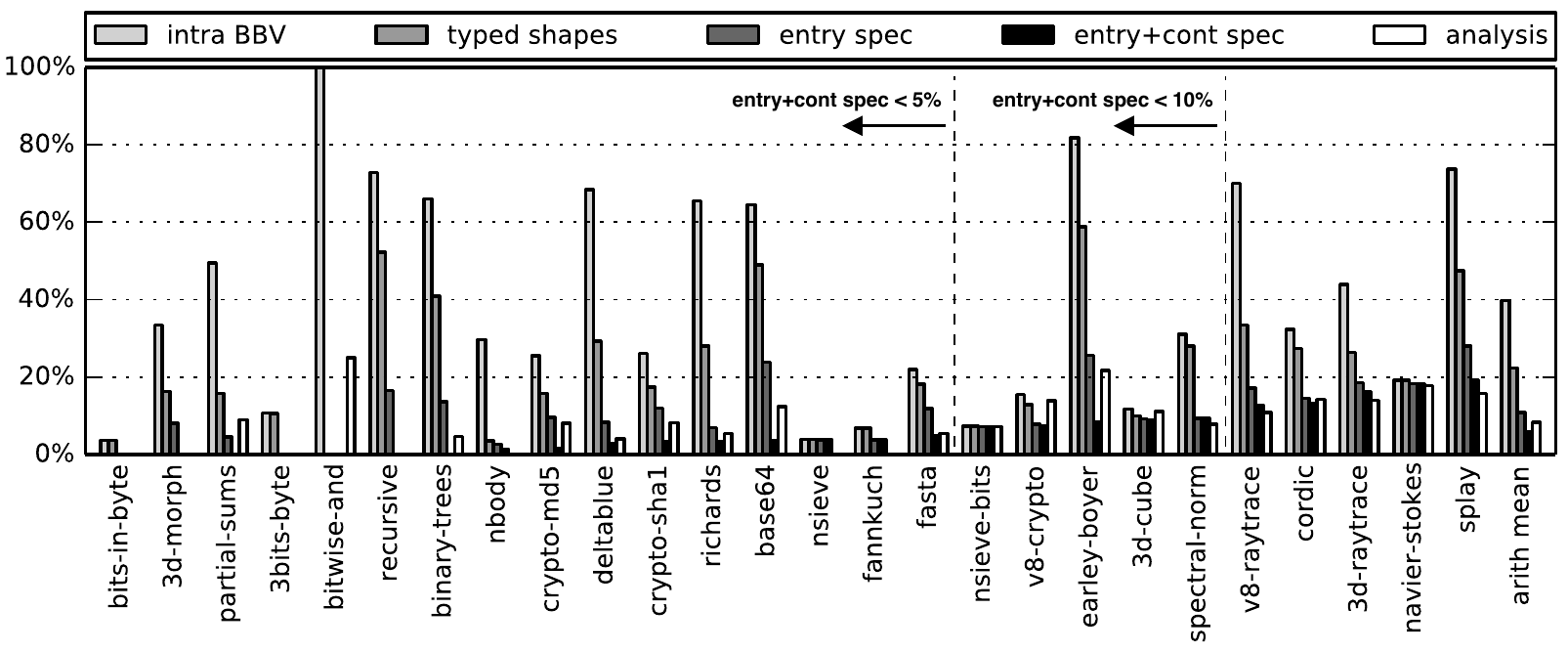}
    \end{center}
\caption{Proportion of type tests remaining relative to a baseline without BBV (lower is better)\label{fig:num_type_tests}}
\end{figure*}

It is useful to compare the performance of BBV in terms
of type tag tests eliminated against a traditional static type analysis for
JS.
However, implementing such an analysis can be time consuming and error prone.
It also raises the issue as to which specific kind of type analysis should be
implemented, and whether the comparison is fair or not.

In order to avoid these issues, a simulated ``perfect''
analysis was used. Each benchmark was run and the
result of all tag tests was recorded. The benchmarks were then rerun with all type tests
that always evaluate to the same result removed. 
The simulated analysis is essentially an oracle with perfect knowledge of the
future behavior of a program. It knows which type tests are redundant based
on past executions on the same input.
This gives an upper bound on the best performance achievable with a static
type analysis in terms of tag tests eliminated with the constraint that the
code is not transformed (for instance, no code duplication).

Figure~\ref{fig:num_type_tests} shows the relative proportion of type tests
executed with different variants of BBV as well as with
the simulated analysis. The first column shows the type tests remaining with
plain intraprocedural BBV, as introduced in~\cite{bbv_ecoop}. The second column shows
the results obtained by adding typed shapes to intraprocedural BBV, as
in~\cite{typed_shapes}. The third column adds entry point specialization, and
the fourth column finally adds call continuation specialization as introduced
in this paper. Lastly, the fifth column shows the results of the perfect
analysis.

Plain intraprocedural BBV eliminates 60\unskip\%
of type tests on average, and typed shapes brings us to
78\unskip\%. The addition of entry point
specialization improves the result further to
89\unskip\% of type tests eliminated.
Finally, completing our interprocedural implementation of BBV with the
addition of call continuation specialization allows us to reach
94.3\unskip\% of type tests
eliminated, in several cases, nearly 100\%.

It is surprising that interprocedural BBV performs as well or
better than the perfect analysis on most benchmarks. The analysis eliminates
just 91.7\unskip\% of type tag
tests on average, less than what interprocedural BBV achieves.
BBV's ability to duplicate certain code paths allows it to track types more precisely and
eliminate more type tests than is possible without transforming code.

\subsection{Call Continuation Specialization}

As outlined in Section~\ref{sec:cont_spec}, call continuation specialization
uses a speculative strategy to propagate return type information without
dynamic dispatch. That is, call continuations for a given callee function may
be recompiled and deoptimized if values are returned which do not match
previously encountered return types for the said function.

Empirically, across our benchmark sets, 10\unskip\% of
functions compiled cause the invalidation of call continuation code.
Dynamically, the type tag of return values is successfully propagated and
known to the caller 81\unskip\% of the time. In over
half of the benchmarks, the type tag of return values is known over 99\% of the
time.

\subsection{Code Size}

Adding entry point and continuation specialization to our existing BBV
implementation cause an average increase in machine code size of
7.6\unskip\% in the worst case and just
2.2\unskip\% on average. 
Intuitively, one may have assumed that a bigger code size increase might have
occurred, given that entry point versioning can generate multiple entry points
per function. However, better optimized machine code tends to be
more compact. We argue that the slight code size increase is observed is
reasonable, particularly given the performance improvements obtained.

\subsection{Compilation time}

The addition of entry point and continuation specialization to our BBV
implementation cause a compilation time increase of
4.7\unskip\% in the worst case and 
1.0\unskip\% on average.
The increase in compilation time is relatively small, and roughly corresponds
to the increase in code size.

\subsection{Execution Time}

\begin{figure*}[tb]
    \begin{center}
    \includegraphics[scale=1.00]{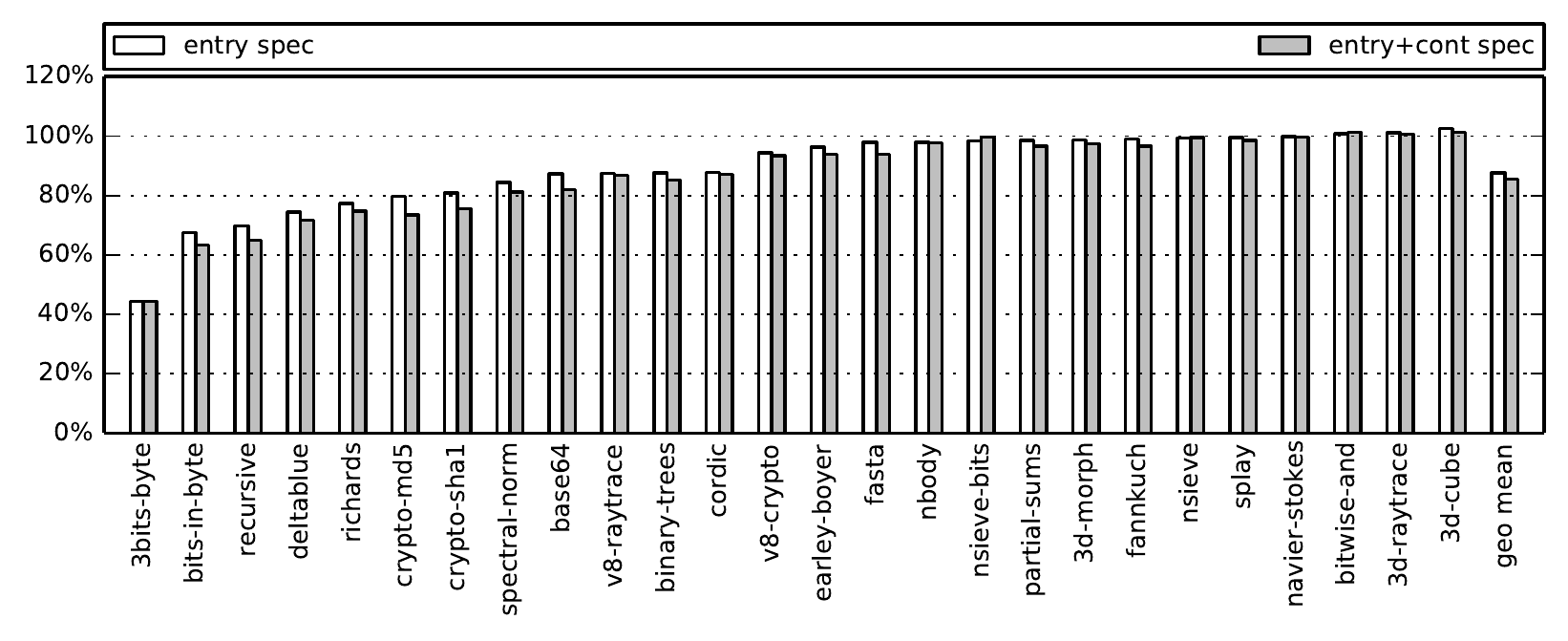}
    \end{center}
\caption{Execution time relative to baseline (lower is better)\label{fig:exec_time}}
\end{figure*}

Figure~\ref{fig:exec_time} shows the relative execution time with entry point
specialization as well as with both entry point and continuation
specialization.
When both entry point and continuation specialization are used,
speedups of 14.5\unskip\% are achieved
on average, and up to 56\unskip\% in
the case of 3bits-bytes.
Importantly, on nearly all benchmarks, the addition of continuation specialization does
improve the average performance over entry point specialization only.

\subsection{Comparison with Truffle/JS}

\begin{figure*}[tb]
    \begin{center}
    \includegraphics[scale=1.00]{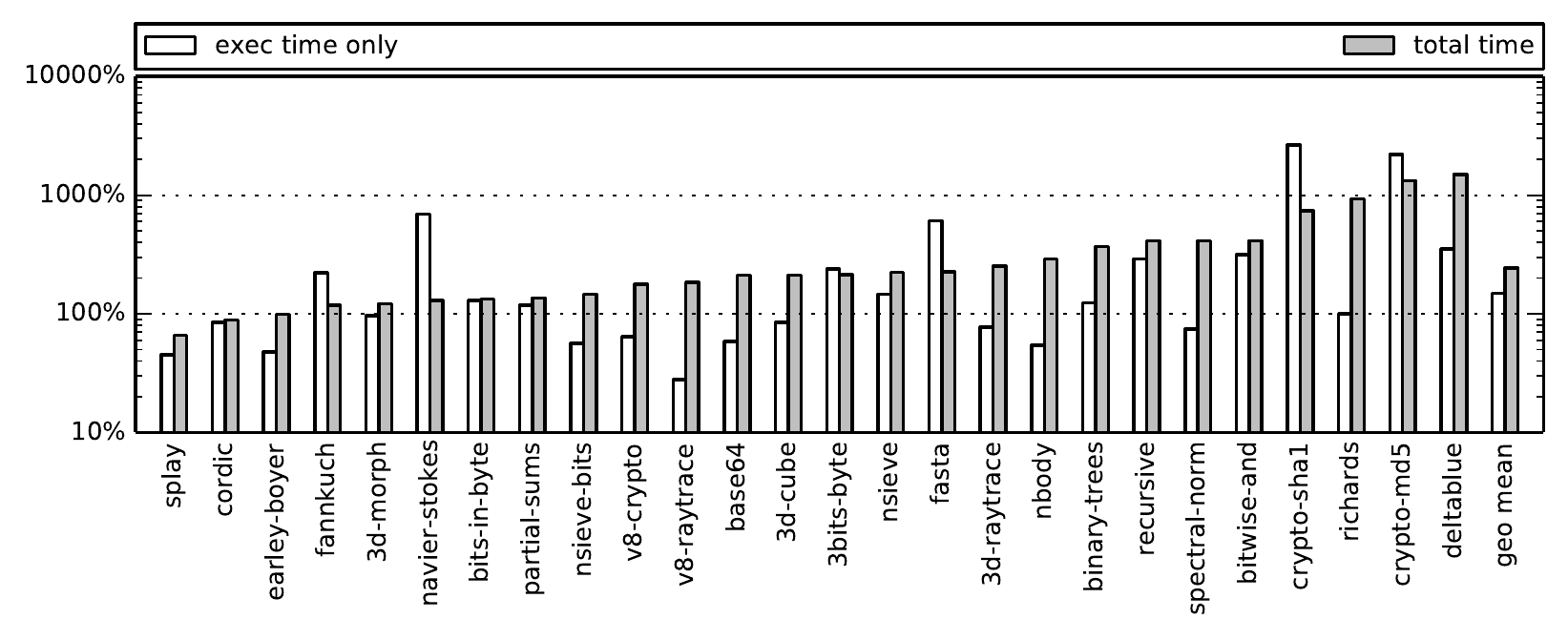}
    \end{center}
\caption{Speedup relative to Truffle (log scale, higher favors Higgs)\label{fig:against_tr}}
\end{figure*}

Truffle/JS, like Higgs, is a research VM written in a garbage-collected language.  For this reason it is interesting to compare their performance.

Since Truffle takes a relatively long time to compile and optimize
code, each benchmark is run for 1000 ``warm up'' iterations before
measuring execution time, so as to more accurately measure the final
performance of the generated machine code once a steady state is reached.
After the warm up period, each benchmark is run for 100 timing iterations.

Figure~\ref{fig:against_tr} shows the results of the performance comparison
against Truffle/JS. The left column represents the speedup of Higgs over Truffle
when taking only execution time into account, warmup iterations excluded.
The right column is the speedup of Higgs over Truffle/JS when comparing
total time, including initialization, compilation and warmup.
A logarithmic scale was used due to the wide spread of data points.

On average, when taking only execution time into account, Higgs outperforms
Truffle on half of the benchmark set, and is
1.49$\times$ 
faster than Truffle/JS on average.
When comparing wall clock times, Higgs outperforms Truffle on the majority of
benchmarks, and is
2.44$\times$ 
faster on
average.

It is not the case that Higgs outperforms Truffle/JS on every benchmark.
JIT compiler design is a complex problem space, and as a result,
there are benchmarks where each system outperforms the other by a wide margin.
Truffle/JS is a method-based compiler implementing profiling, type-feedback
and type analysis. It has a few technical advantages over Higgs,
such as the use of method inlining, a more sophisticated register allocator,
multithreaded background compilation, and a highly optimized garbage collector
implementation (the Java VM's). These tools give Truffle an edge on
specific benchmarks, such as {\tt v8-raytrace}. Nevertheless, Higgs
produces competitive machine code in many cases.
It is interesting to see that using the simpler interprocedural BBV approach
yields faster compilation (no warmup to speak of), and faster code on
average.  As a rough measure of compiler complexity, the source code of
Truffle/JS is about 6 times larger than Higgs'.

\section{Limitations and Future Work}\label{sec:future}
While the results obtained with interprocedural BBV are in many cases
superior to what is achievable with a traditional type analysis, there is still
room for improvement. Our system does not keep
track of the types of captured closure variables. It would be desirable to
further extend BBV to do so. The system also treats
arrays like untyped black boxes. Extending the system to keep track of the
types of values stored inside of arrays, in particular numerical types, would
likely lead to performance improvements.



\section{Related Work}

\subsection{Type Analysis of Dynamic Languages}

Basic block versioning bears resemblance to the {\em iterative type analysis} and
{\em extended message splitting} techniques developed for Self by Craig
Chambers and David
Ungar~\cite{itr_analysis}. This is a combined static analysis and
transformation that compiles multiple versions of loops and duplicates 
control flow paths to eliminate type tests. The analysis works in an iterative
fashion, transforming the control flow graph of a function while performing
a type analysis. It integrates a mechanism to generate new versions of loops
when needed, and a message splitting algorithm to try and minimize type
information lost through control flow merges. One key disadvantage
is that statically cloning code requires being conservative,
generating potentially more code than necessary, as it is impossible to
statically determine exactly which control flow paths will be taken at run
time, and this must be overapproximated. Basic block versioning is simpler to
implement and generates code lazily, requiring less compilation time and memory
overhead, making it more suitable for integration into a baseline JIT compiler.

There have been multiple efforts to devise type analyses for dynamic languages.
The Rapid Atomic Type Analysis (RATA)~\cite{rata} is an intraprocedural
flow-sensitive analysis based on abstract interpretation that aims to assign
unique types to each variable inside of a function. Attempts have
also been made to define formal semantics for a subset of dynamic languages
such as JavaScript~\cite{ti_js}, Ruby~\cite{ti_ruby} and Python~\cite{rpython},
sidestepping some of the complexity of these languages and making them more
amenable to traditional type inference techniques. There are
also flow-based interprocedural type analyses for JavaScript based on
sophisticated type lattices~\cite{tajs}\cite{tajs_lazy}\cite{type_ref}. Such
analyses are usable in the context of static code analysis, but take too long
to execute to be usable in VMs and do not deal with the complexities of
dynamic code loading.

The work done by Kedlaya, Roesch et al.~\cite{impr_type_spec} shows strategies
for improving the precision of type analyses by combining them with type
feedback and profiling. This strategy shows promise, but does not explicitly
deal with object shapes and property types. Work has also been done on a
flow-sensitive alias analysis for dynamic languages~\cite{alias_dynamic}. This
analysis tries to strike a balance between precision and speed, it is likely
too expensive for use in JIT compilers, however.

More recently, work done by Brian Hackett et al. at Mozilla resulted in an
interprocedural hybrid type analysis for JavaScript suitable for use in
production JIT compilers~\cite{mozti}. This analysis represents an important step
forward for dynamic languages, but as with other type analyses, must
conservatively assign one type to each value, making it vulnerable to
imprecise type information polluting analysis results. Basic block versioning
can help improve on the results of such an analysis by hoisting
tests out of loops and generating multiple optimized code paths where
appropriate.

\subsection{Trace Compilation}

{\em Trace compilation}, originally introduced by the
Dynamo~\cite{dynamo} native code optimization system, and later applied to
JIT compilation in HotpathVM~\cite{hotpathvm} aims to record long sequences
of instructions executed inside hot loops. Such linear sequences of
instructions often make optimization simpler. Type information can be
accumulated along traces and used to specialize code and remove type
tests~\cite{trace_type_spec}, overflow checks~\cite{trace_ovf} or unnecessary
allocations~\cite{trace_alloc}. Basic block versioning resembles tracing in
that context updating works on essentially linear code fragments and code is
optimized similarly to what may be done in a tracing JIT. Code is also
compiled lazily, as needed, without compiling whole functions at once.
Trace compilation~\cite{pypy} and meta-tracing are still an active area of
research~\cite{bolz_tracing_racket}.

The simplicity of basic block versioning is one of its main advantages.
It does not require external infrastructure such as an
interpreter to execute code or record traces. Trace compiler
implementations must deal with corner cases that do not appear with
basic block versioning. With trace compilation, there is the potential for
trace explosion if there is a large number of control flow paths going through
a loop. It is also not obvious how many times a loop should be recorded or
unrolled to maximize the elimination of type checks. This problem is
solved with basic block versioning since versioning is driven
by type information. Trace compilers must implement parameterizable policies
and mechanisms to deal with recursion, nested loops and potentially very long
traces that do not fit in instruction caches.

\subsection{Just-In-Time Code Specialization}

{\em Customization} is a technique developed to optimize Self
programs~\cite{self_customization} that compiles multiple copies of methods
specialized on the receiver object type. Similarly, {\em type-directed
cloning}~\cite{type_cloning} clones methods based on argument types,
producing more specialized code using richer type information. The work of 
Chevalier-Boisvert et al. on {\em Just-in-time specialization} for
MATLAB~\cite{mcvm} and similar work done for the MaJIC MATLAB
compiler~\cite{majic_matlab} tries to capture argument types to dynamically
compile optimized versions of whole functions. All of these techniques are
forms of type-driven code duplication aimed at extracting type information.
Basic block versioning operates at a lower level of granularity, allowing
it to find optimization opportunities inside of method bodies by duplicating
code paths.

There are notable similarities between the Psyco JIT specialization work and
our own. The Psyco prototype for Python~\cite{psyco} is able to interleave
execution and JIT compilation to gather run time information about values, so as
to specialize code on-the-fly based on types and values. It also incorporates
a scheme where functions can have multiple entry points. 
We extend upon this work by combining a similar approach, that of basic block
versioning, with typed shapes and a mechanism for propagating return types
from callees to callers with low overhead.

The {\em tracelet-based} approach used by Facebook's {\em HHVM} for
PHP~\cite{hiphopvm} bears
important similarities to our own. It is based on the JIT compilation of small code
regions (tracelets) which are single-entry multiple-exit basic blocks. Each
tracelet is type-specialized based on variable types observed at JIT
compilation time. Guards are inserted at the entry of tracelets to verify at
run time that the types observed are still valid for all future executions.
High-level instructions in tracelets are specialized based on the guarded
types. If these guards fail, new versions of tracelets are compiled based
on different type assumptions and chained to the failing guards.

There are three important differences between the HHVM approach and basic block
versioning. The first is that BBV does not insert dynamic guards but instead
exposes and exploits the underlying type checks that are part of the definition
of runtime primitives. HHVM cannot do this as it uses monolithic high-level
instructions to represent PHP primitives, whereas the Higgs primitives are
self-hosted and defined in an extended JavaScript dialect.
The second difference is that BBV propagates known types to successors and 
doesn't usually need to re-check the types of local variables. Finally,
HHVM uses an interpreter as fallback when too many tracelet versions
are generated. Higgs falls back to generic basic block versions which do not
make type assumptions but are still always JIT compiled for better performance.

\section{Conclusion}\label{sec:conclusion}
Basic Block Versioning (BBV) is a JIT compilation strategy for generating
type-specialized machine code on-the-fly without a separate type analysis
pass. We have shown that it
can be successfully extended with techniques to propagate information across
method call boundaries, both from callers to callees and from callees to
callers, and this without requiring dynamic dispatch.

Across 26\unskip benchmarks, interprocedural BBV
eliminates 94.3\unskip\% of type tests
on average, more than a simulated static type analysis with access to perfect
information. The addition of interprocedural capabilities
to BBV provides an additional
speedup of 14.5\unskip\% on average over an
unextended BBV implementation.


\acks

Special thanks go to Laurie Hendren, Erick Lavoie, Vincent Foley, Paul Khuong,
Molly Everett, Brett Fraley and all those who have contributed to the
development of Higgs.

This work was supported, in part, by the Natural Sciences and Engineering
Research Council of Canada (NSERC) and Mozilla Corporation.

\bibliographystyle{plain}
\bibliography{main}

\end{document}